\documentstyle[epsfig,aps,twocolumn]{revtex}

\newcommand{\rv}{{\bf r}}

\newcommand{\beq}{\begin{equation}}
\newcommand{\eeq}{\end{equation}}
\newcommand{\bea}{\begin{eqnarray}}
\newcommand{\eea}{\end{eqnarray}}

\begin{document}

\draft
\preprint{}
\title{Interference of a Bose-Einstein condensate in a 
hard-wall trap:\\ Formation of vorticity}
\author{J. Ruostekoski, B. Kneer, and W. P. Schleich} 
\address{Abteilung f\"ur Quantenphysik,
Universit\"at Ulm, D-89069 Ulm, Germany}
\date{\today}
\maketitle
\begin{abstract}
We theoretically study the coherent expansion of a Bose-Einstein condensate
in the presence of a confining impenetrable hard-wall potential. 
The nonlinear dynamics of the macroscopically coherent matter 
field results in rich and complex spatio-temporal interference 
patterns demonstrating the formation 
of vorticity and solitonlike structures, and the fragmentation 
of the condensate into coherently coupled pieces.

\end{abstract} 

A landmark experiment \cite{AND97} demonstrated in a striking way
the interference of two freely expanding Bose-Einstein condensates
(BECs). 
In this paper we theoretically study the evolution of a BEC in 
a coherently reflecting hard-wall trap. The present situation
is closely related to the recent experiments on an 
expanding BEC in an optically-induced `box' 
by Ertmer and coworkers \cite{BON99}. Due to
the macroscopic quantum coherence the BEC exhibits rich and complex 
self-interference patterns. 
We identify the formation of vorticity and 
solitonlike structures, and the dramatic fragmentation of an initially
uniform parabolic BEC into coherently coupled pieces.

Atomic BECs exhibit a macroscopic quantum 
coherence in an analogy to the optical coherence of lasers. 
In the conventional reasoning the coherence of a BEC 
is introduced in the spontaneous symmetry breaking. 
Nevertheless, even two BECs 
with no phase information, could show relative phase 
correlations 
as a result of the back-action of quantum measurement \cite{JAV96}.
Moreover, the density-dependent self-interaction of a BEC demonstrates 
the analogy between nonlinear laser optics and {\it nonlinear 
atom optics} \cite{LEN93,DEN99} with BECs. 
BECs are predicted to exhibit dramatic coherence properties:
The macroscopic coherent quantum tunneling \cite{JAV86}
and the formation of fundamental 
structures, e.g., vortices \cite{DAL96,JAC98,BOL98b} 
and solitons \cite{LEN93,REI97}.
Some basic properties of grey 
solitons have been recently addressed for 
harmonically trapped 1D BECs in Ref.~\cite{REI97}. 
Also optical solitons have been actively 
studied in the 1D homogeneous space \cite{KIV98}. 

In this paper we study the dynamics of a BEC confined to a 
hard-wall trap with potential $V(\rv)$. Such walls can be realized,
e.g., with a blue-detuned far-off-resonant light sheet.
Throughout the paper we focus on repulsive interactions. 
When the BEC is released
from a magneto-optical trap (MOT) inside the confining 
potential by suddenly turning off the MOT,
the repulsive mean-field energy of 
the condensate transforms into kinetic energy and the BEC 
rapidly expands towards the walls. 
The reflections of the matter wave from the binding potential result 
in rich and complex spatio-temporal and interference patterns
referred to in 1D as quantum carpets 
\cite{KAP98}. They have been recently proposed as a thermometer for
measuring the temperature of the BEC \cite{CHO99}. In this paper 
we show that the nonlinear dynamics of a BEC 
displays dramatic {\it local}
variations of the condensate phase including the formation of
vorticity and solitonlike structures. The solitary waves could 
possibly be used as
an experimental realization of the macroscopic coherent tunneling
analogous to the Josephson effect.

The dynamics of a BEC follows from the Gross-Pitaevskii equation (GPE) 
$$
i \hbar {\partial\over\partial t}\psi(\rv;t)=\left[ -{\hbar^2\over 2M}
{\bbox \nabla}^2+V(\rv)+\kappa |\psi(\rv;t)|^2 \right]
\psi(\rv;t)\,.
$$
Here $M$ and $\kappa\equiv 4\pi\hbar^2 aN/M$ denote the atomic mass and
the coefficient of the nonlinearity, respectively, with
the scattering length $a$ and the number $N$ of BEC atoms.
Our initial distribution
$\psi(\rv;t=0)$ is the stationary solution of the GPE with
the potential $V(\rv)$ replaced by the potential of
the MOT.

We integrate GPE in one and two spatial dimensions.
The projections from 3D into 1D or 2D require that the mean field
$\psi$ does not vary significantly as a function of time 
in the corresponding orthogonal directions. This condition can be satisfied,
e.g., in the presence of a strong spatial confinement to these dimensions.
Then we can approximate the position dependence in these
directions  by constants ${\cal A}$ and $\ell$ resulting in
$\psi(\rv)\simeq \psi_1(x)/{\cal A}^{1/2}$ in 1D and 
$\psi(\rv)\simeq \psi_2(x,y)/\ell^{1/2}$
in 2D. This yields 
the strengths $\kappa_1=\kappa/{\cal A}$ and $\kappa_2=\kappa/\ell$
of the nonlinearity in GPE for the mean fields in 1D and 2D
$\psi_1$ and $\psi_2$.
We emphasize that especially the 2D calculations may already
contain the essential features of the full 3D 
coupling between the different spatial dimensions
by the nonlinearity.

{\it One-dimensional case --} The linear Schr\"odinger equation for the
1D box of length $L$
exhibits \cite{KAP98} regular spatio-temporal patterns. These patterns
consist of
straight lines, so called traces, of different steepness corresponding 
to harmonics of a fundamental velocity, $v_0$. The traces
arise from the interferences between degenerate eigenmodes of 
the system. 
The eigenmodes of frequency $\omega_n\equiv n^2\omega_1
\equiv n^2\hbar\pi^2/(2ML^2)$ 
are a superposition of right and left 
propagating plane waves with wavenumbers $k_n\equiv n\pi/L
\equiv n k_1$. The 
probability density consists of the interferences between
different eigenmodes. Therefore the lines of constant phase
$\pm (k_m\pm k_n)x+(\omega_m-\omega_n)t={\rm const}$. 
correspond to straight lines in space-time with velocities
$v_{mn}\equiv \pm (\omega_m-\omega_n)/(k_m\pm k_n)=
(m\pm n)v_0$. Here we introduced the fundamental trace velocity
$v_0\equiv \hbar\pi/(2mL)$.

\begin{figure}
\begin{center}
\leavevmode
\begin{minipage}{4.2cm}
\epsfig{
width=4.2cm,file=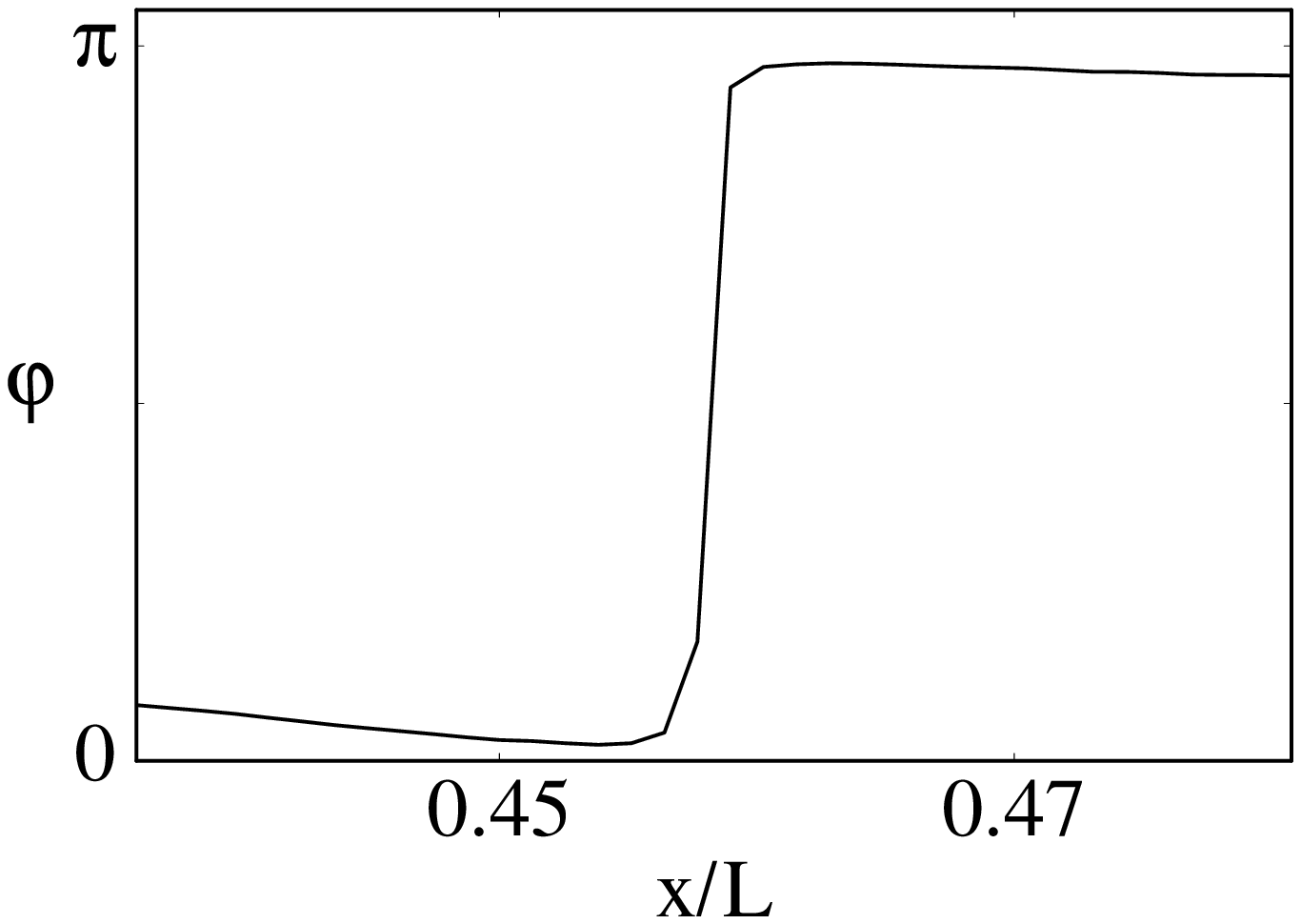}
\end{minipage}
\begin{minipage}{4.2cm}
\epsfig{
width=4.2cm,file=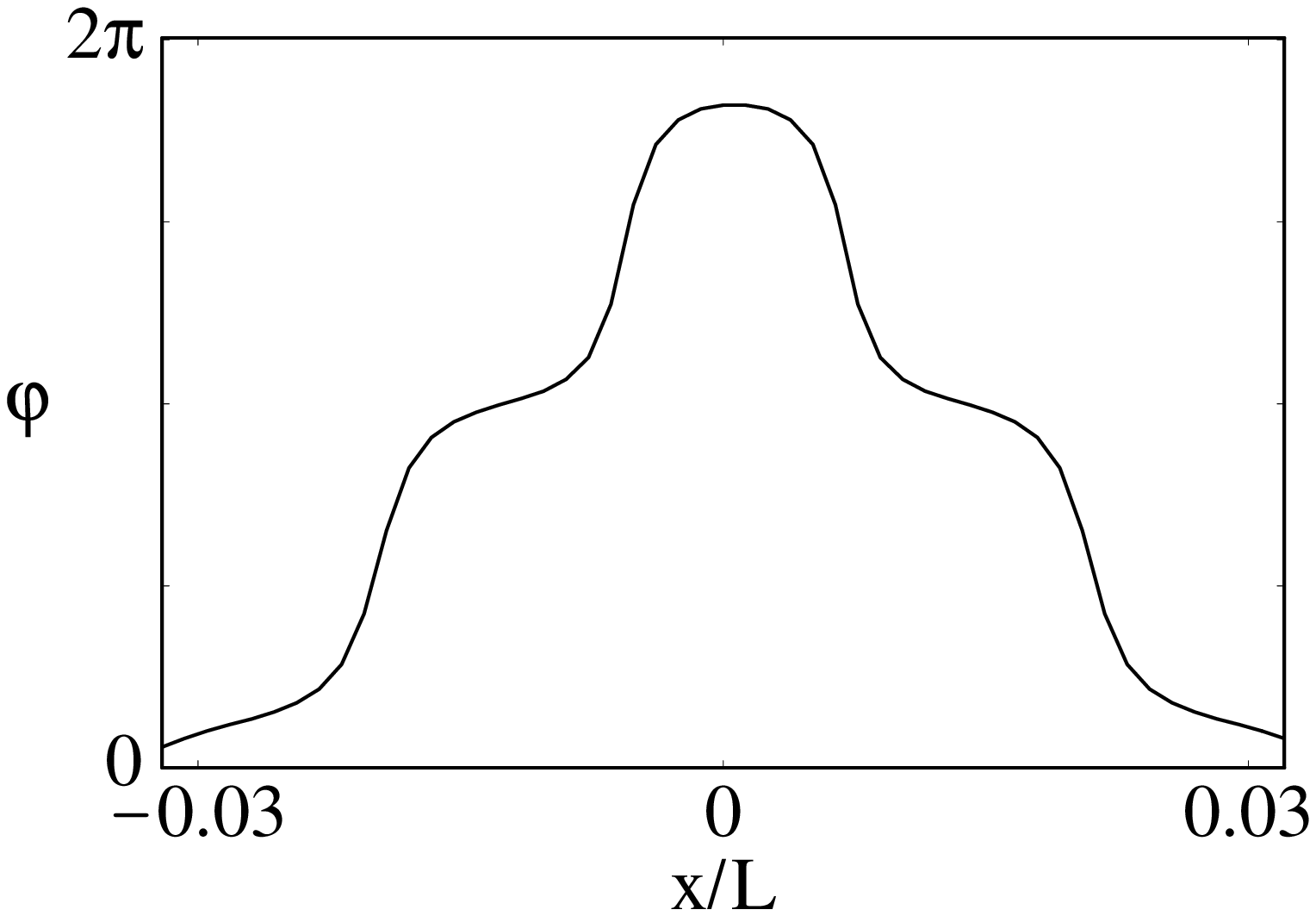}
\end{minipage}
\end{center}
\caption{
[Part (a) in a separate file]
Quantum carpet of a BEC woven by propagating grey solitonlike structures.
Time evolution (a) of a BEC density profile $|\psi_1(x;t)|^2$
in a 1D hard-wall trap. 
The horizontal and vertical 
axes denote the spatial coordinate and time, respectively.
The initially confined BEC expands towards, and reflects from,
the hard-wall boundaries. The dark canals correspond to low
density and represent the evolution of grey solitonlike structures. 
The phase slip of a single fringe (b) at $t/T\simeq 3.2\times10^{-3}$, and
of four colliding fringes (c) at $t/T\simeq 0.021$.
The arrows in (a) denote the times of the phase profiles (b) and (c).
Here time is scaled in units of 
the revival time $T\equiv 2\pi/ \omega_1\equiv 4ML^2/(\pi\hbar)$
of the linear Schr\"odinger equation and
the initial state is $R/L\simeq0.3$ and $\kappa_1=19000\hbar/T$. 
}
\label{fig:1}
\end{figure}

We now turn to the quantum carpet of Fig.~\ref{fig:1}a representing
a 1D BEC trapped between two impenetrable steep Gaussian potentials
approximating infinitely high walls at $x=\pm L/2$.
At time $t=0$ the BEC is released from a harmonic 
trap of frequency $\Omega$. In the limit of strong confinement
the initial state is well 
approximated by the Thomas-Fermi solution
$\psi_1(x;t=0)=\theta(R_1-|x|)[3(R_1^2-x^2)/(2R_1^3)]^{1/2}$. Here
$R_1\equiv [3\kappa_1/(M\Omega^2)]^{1/3}$ 
describes the 1D radius of the BEC wave function. 

After the turn-off of the MOT
the kinetic energy term in GPE becomes dominant
and the matter wave expands towards, and eventually reflects from, the 
box boundaries. Due to the macroscopic quantum coherence 
of a BEC different spatial regions of the matter field generate a complex
self-interference pattern that exhibits canals analogously
to the quantum carpet structures of the linear Schr\"odinger equation 
\cite{KAP98}. 
From Fig.~\ref{fig:1} we note that only destructive interference fringes,
canals, appear in the nonlinear carpet. Constructive interference
fringes, ridges, do not emerge.

For the single particle Schr\"odinger equation
the resulting quantum carpet
demonstrates the fundamental {\it wave nature} of the particle.
Therefore, it is perhaps surprising that in the case of GPE, which represents
the coherent matter {\it field} of the many-particle system, the intermode
traces acquire properties that demonstrate dramatic {\it particle
nature}. In particular, in the case of repulsive 
interparticle interactions the traces correspond to the evolution 
of solitonlike structures with an associated phase kink 
in Fig.~\ref{fig:1}b,c.

Grey solitons correspond to the propagation of `density holes' in 
the matter field and the depth of the hole characterizes the greyness 
of the soliton. In homogeneous space the soliton is called dark,
when it exhibits \cite{KIV98,REI97} a vanishing 
density at the center of the dip and a sharp phase slip of $\pi$.
For grey solitons the value of the phase slip $\varphi$ is reduced, thereby 
giving the soliton a nonvanishing speed according to 
$v=c\cos(\varphi/2)$, where the maximum velocity
$c\equiv (4\pi\hbar^2a\rho/M^2)^{1/2}$ 
is the Bogoliubov speed of sound at a constant atom density $\rho$.
As a result of the formation of the soliton-like structures the initially
uniform parabolic BEC is dramatically fragmented into spatially 
separated pieces and
we can interprete these density holes  as the 
boundaries of the fragmented BEC. Analogously to the Josephson junction
the phase of the BEC is approximately constant outside the narrow region
of the solitary wave. The coherent tunneling of atoms across the boundary
depends on the relative phase between the two contiguous pieces
and results in the motion of the `solitary wave junction'. 
For a phase slip of $\pi$
the velocity of the soliton vanishes analogously to the vanishing
Josephson oscillations of the number of atoms.

The interatomic interactions shift the eigenenergies 
compared to the linear case. As a result the fringe velocities 
in Fig. \ref{fig:1} are a few tens of percents larger than 
the linear trace velocities $4nv_0$ for the canals with 
a symmetric initial state \cite{KAP98}. The nonlinear fringe velocities
are still determined by the degenerate intermode traces and are 
approximately integer multiples of the minimum speed. The largest 
velocity in Fig.~\ref{fig:1} is approximately given by the
Bogoliubov speed of sound in terms of the average density of atoms
in the box. 

The fringe evolution 
shown in Fig.~\ref{fig:1} displays a remarkable
inherent particle-like robustness and solitonlike behaviour: 
The fringes survive 
complex dynamics. Their paths exhibit dramatic avoided 
crossings demonstrating repulsive interactions and therefore
the colliding wave packet holes become degenerate in momentum space
\cite{RUO99d}. 

{\it Two-dimensional case --} We now turn to the evolution
of a BEC in a circular box. Again we approximate the infinitely
high walls by a steep Gaussian potential aligned
along a circle of radius $r\equiv L/2$. 
The initial wave function is the Thomas-Fermi solution of the
symmetric MOT
$\psi_2(x,y;t=0)=[2(R_2^2-x^2-y^2)/(\pi R_2^4)]^{1/2}$, and
$R_2\equiv [4\kappa_2/(\pi M\Omega^2)]^{1/4}$ denotes the 2D
`classical' radius. Hence, the initial wave function is located
at the center of the circular box and we can show that in this 
case the state remains symmetric also at later times.

\begin{figure}
\begin{center}
\leavevmode
\begin{minipage}{4.2cm}
\epsfig{
width=4.2cm,file=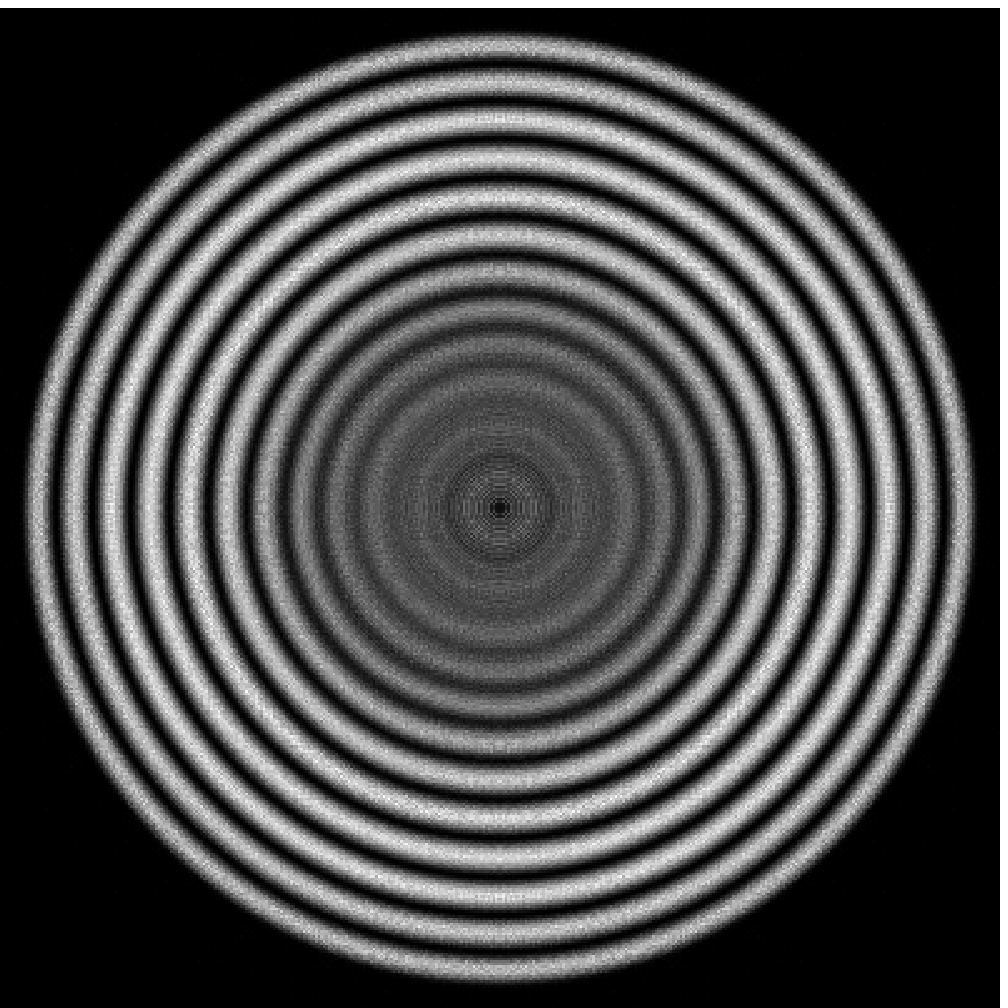}
\end{minipage}
\begin{minipage}{4.2cm}
\epsfig{
width=4.2cm,file=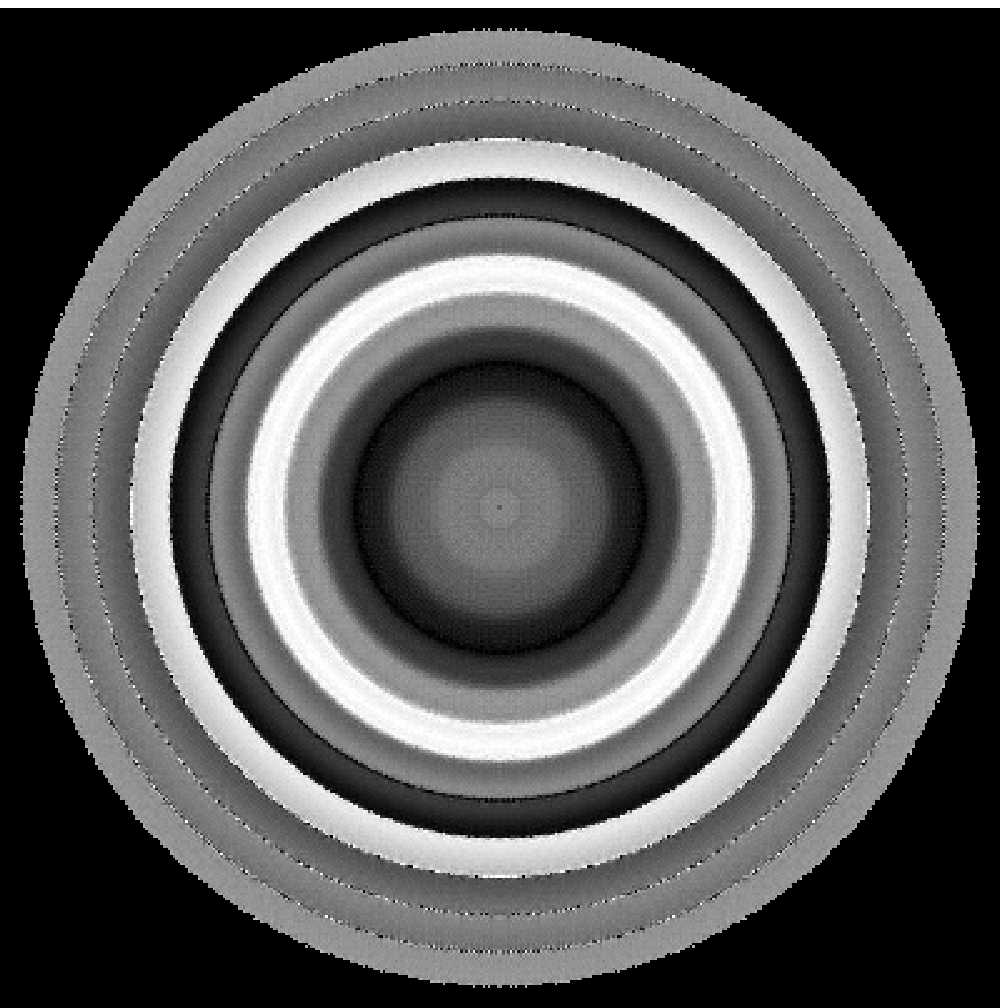}
\end{minipage}
\end{center}
\caption{
Formation of loops of solitonlike structures
in a BEC expanding in a 2D circular box.
The symmetric condensate has started from the center of the circular box,
has expanded and reflected from the boundary.
We show the density (left) and phase (right) profiles of the BEC at time 
$t=4.9\times10^{-3}ML^2/\hbar$.
In the phase profile we have chosen
a continuous phase $|\varphi|$, where $-\pi\leq \varphi\leq \pi$.
The initial state is $R_2/L\simeq0.28$ and $\kappa_2\equiv 1500 \hbar^2
/(ML^2)$.
}
\label{fig:2}
\end{figure}

Figure \ref{fig:2} shows the 2D density $|\psi_2(x,y;t)|^2$
and the phase profile $|\varphi|$ of a BEC obtained from GPE
at a later time.
We note the formation of a regular interference patterns 
similar to solitary waves.
The fragmentation of the BEC is even more dramatic than in the 1D case:
The fringes exhibit a vanishing density at the center of the dip.
The BEC forms coherently coupled loops and the
resulting structures are similar to optical ring solitons~\cite{KIV98}.

When we now slightly displace the initial state of the BEC
from the center of the circular box,
the rotational symmetry is broken.  In Figs. \ref{fig:3} and  \ref{fig:3b}
we show the resulting 2D density (left column) and the phase profiles 
(right column) at three characteristic times.
The reflections of the BEC
from the hard-wall potential create solitonlike structures. At later times
the stripes bend and eventually break up
forming dark spots that correspond to vortices 
with associated phase windings around closed paths \cite{DAL96}.

We note that we can use the present situation of an expanding
BEC in a circular box to create vortices in some particular spatial 
location by simply introducing a {\it static} potential dip 
and letting the expanding BEC flow across it. This is a simplified
version of the suggestion by 
Ref.~\cite{JAC98} that a moving potential barrier
through a BEC can create vorticity in the vicinity of the potential.

As a final example and to demonstrate the effect of the symmetry
of the hard-wall
trap we consider the evolution of a BEC in the 2D square
box. We generate the boundary by steep Gaussians approximating infinitely
high walls at $x=\pm L/2$ and $y=\pm L/2$. 

\begin{figure}
\begin{center}
\leavevmode
\begin{minipage}{4.2cm}
\epsfig{
width=4.2cm,file=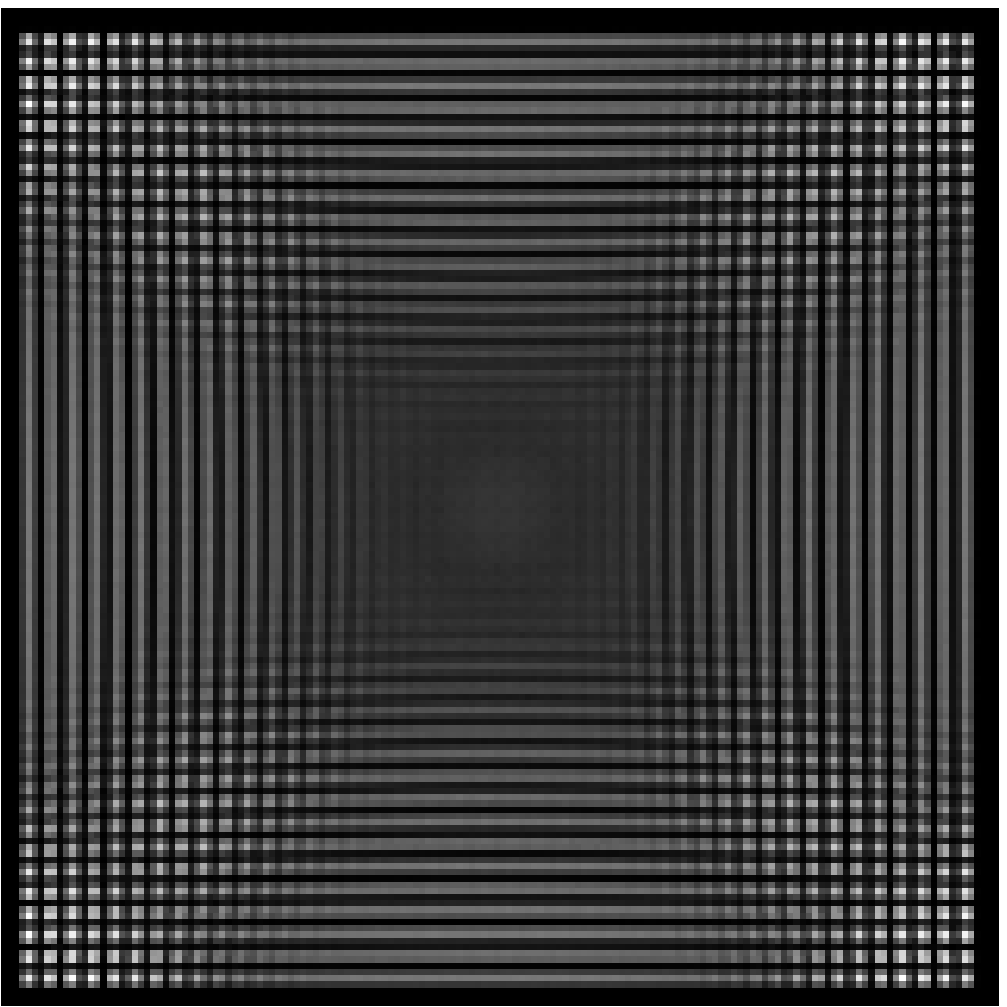}
\end{minipage}
\begin{minipage}{4.2cm}
\epsfig{
width=4.2cm,file=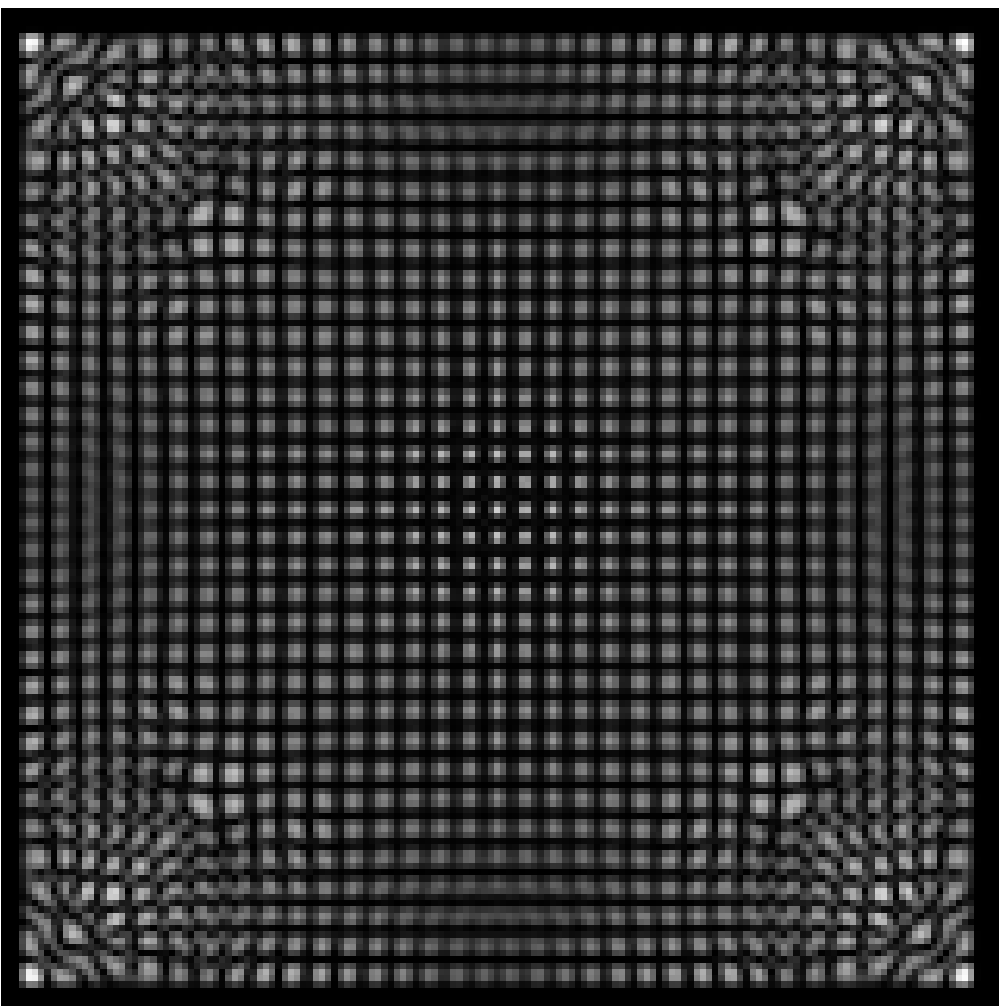}
\end{minipage}
\end{center}
\caption{
Expanding BEC in a 2D square box with
a symmetric initial state at the center of the square.
At $t=2.9\times10^{-3}ML^2/\hbar$ (left) the density profile of the
BEC exhibits a rectangular pattern of solitonlike structures.
At a later time $t=3.9\times10^{-3}ML^2/\hbar$ (right) 
the pattern is distorted and we see the beginning 
of the formation of vorticity symmetrically around the diagonals.
The initial radius of the
BEC is $R_2/L\simeq0.28$ and $\kappa_2=5000\hbar^2/(ML^2)$.
}
\label{fig:4}
\end{figure}

The square has the symmetry of rotations of $\pi/2$. Hence, for a
symmetric nonrotating initial state the minimum number of vortices 
conserving the total angular momentum is eight. 
In Fig.~\ref{fig:4}
we show the density profile at two characteristic times. The
reflections of the BEC from the boundary generate dark solitonlike
structures with an amazingly regular square shape that
start bending, break up, and form vorticity.

We now address the experimental feasibility
of the proposed self-interference measurements.
A higher-order Laguerre-Gaussian beam can 
generate a hollow optical beam 
\cite{MAN98} and thus
a cylindrical atom-optical hard-wall potential around the 
magnetically-trapped BEC.
Moreover, BECs with purely 2D confinement may 
also be investigated in
certain magnetic trap configurations \cite{ERN98}.
In recent experiments Bongs {\it et al.} \cite{BON99} 
studied the coherent reflections of a BEC by atom-optical mirrors
and the evolution of a BEC in a atom-optical waveguide closely
related to the present theoretical study.

\begin{figure}
\caption{
[In a separate file]
Formation of vorticity in a BEC expanding in a 2D circular box with
displaced initial state.
We show the density profiles (left column) and the corresponding
phase profiles (right column) of the BEC at three 
characteristic times $t=2.5\times10^{-3}$, $4.9\times10^{-3}$,
and $6.1\times10^{-3}$ in units of $ML^2/\hbar$. 
The reflections from the boundary generate
solitonlike structures that bend, break up, and form vorticity.
The velocity of the BEC is proportional to the
gradient of the phase. 
}
\label{fig:3}
\end{figure}

The density profiles of the BECs could be directly measured via
absorption imaging, if the necessary spatial resolution could
be obtained, e.g., via ballistic expansion of the atomic cloud.
Vortices may also be detected by interfering a BEC
with and without vorticity. Then the phase slip in the
interference fringes would be the signature of the vorticity \cite{BOL98b}.
The phase slip between two pieces of the BEC has a dramatic 
effect on the dynamical structure factor of the two-component 
system~\cite{RUO97b} which may be observed, e.g., via the Bragg 
scattering~\cite{STA99}.

In conclusion, we studied the generation of vorticity and 
solitonlike structures of a BEC in a hard-wall trap. 
The nonlinear evolution of GPE dramatically divides the
initially uniform parabolic BEC into coherently coupled pieces. 
We showed that the density profile of the BEC
can be a direct manifestation of the macroscopic
quantum coherence. Obviously it could also be a sensitive measure 
for the decoherence rate of the BEC \cite{HAL98b}. Unlike the typical 
coherence measurement that detects the relative macroscopic phase
between two well-distinguishable BECs \cite{AND97,HAL98b}, the present
set-up probes the self-interference of an initially uniform matter field.

We acknowledge financial support from DFG and from the
EC through the TMR Network ERBFMRXCT96-0066.

\begin{figure}
\begin{center}
\leavevmode
\begin{minipage}{4.06cm}
\epsfig{
width=4.06cm,file=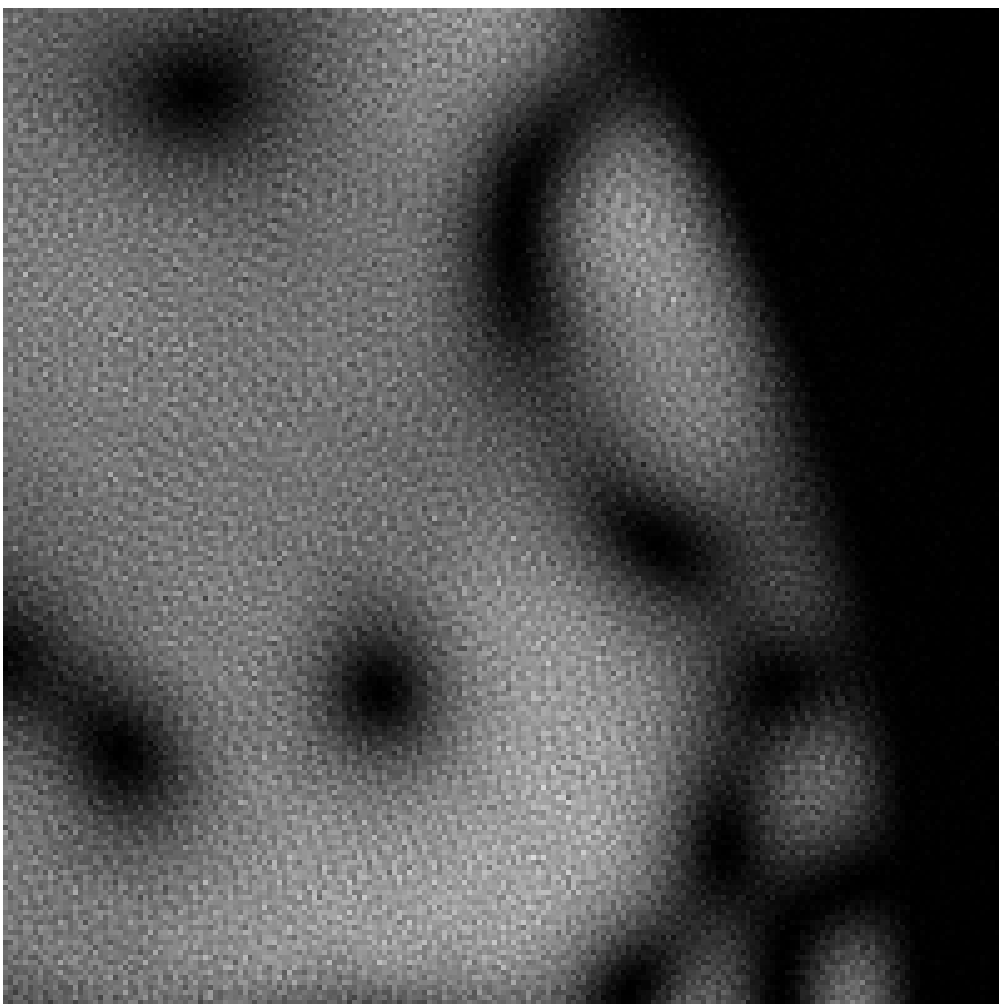}
\end{minipage}
\begin{minipage}{4.2cm}
\epsfig{
width=4.2cm,file=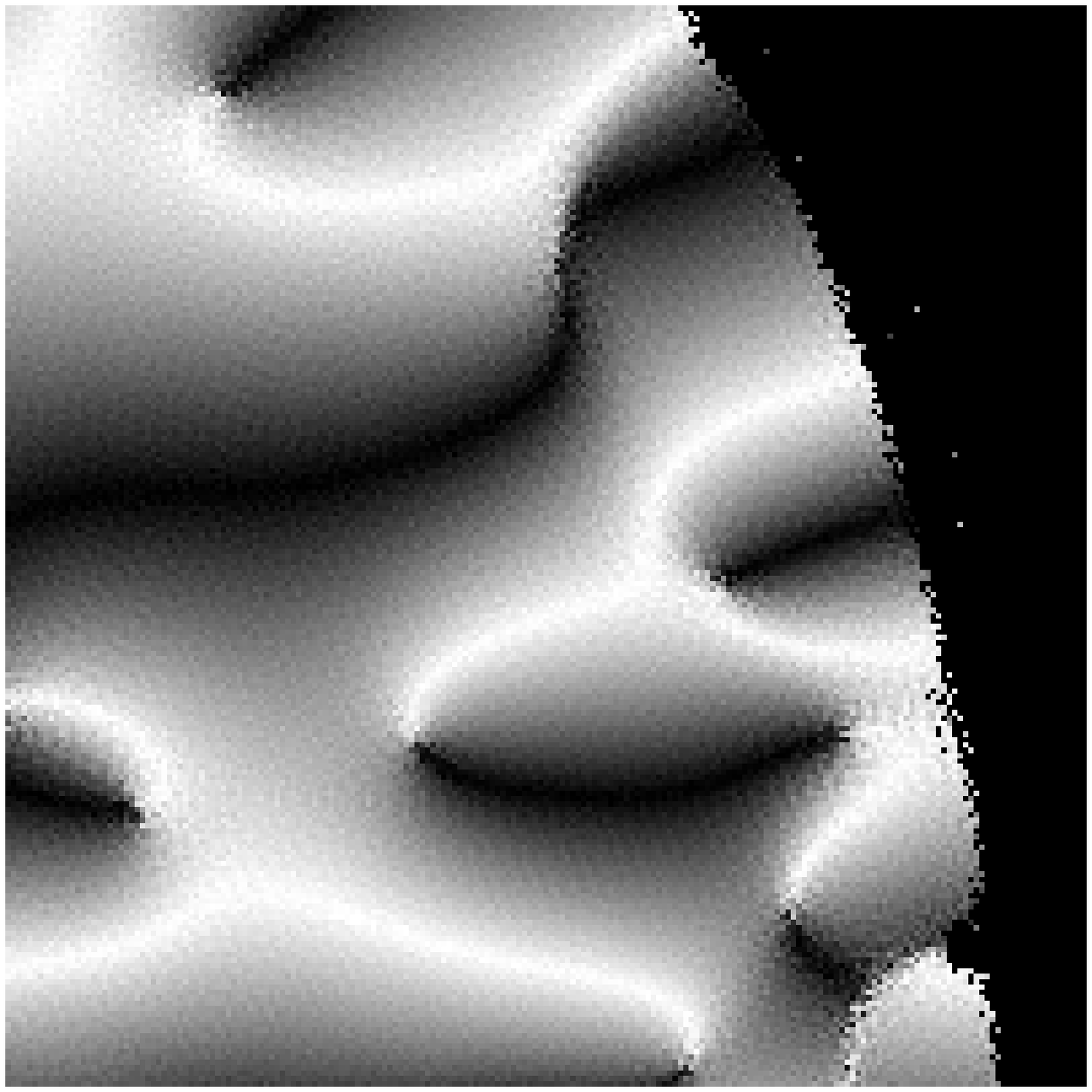}
\end{minipage}
\end{center}
\caption{
The magnification of the last density and phase profile of 
Fig. \ref{fig:3} displaying vorticity. For example, the dark
spot in the center represents a vortex with a unit circular
quantization of $2\pi$.
}
\label{fig:3b}
\end{figure}

\begin{references}

\bibitem{AND97} M.R. Andrews {\it et al.}, Science {\bf 275}, 637 (1997).

\bibitem{BON99} K. Bongs {\it et al.}, unpublished.

\bibitem{JAV96} J. Javanainen and S.M. Yoo, Phys. Rev. Lett. {\bf 76}, 161
(1996); S.M. Yoo {\it et al.}, J. Mod. Opt. {\bf 44}, 1763 (1997).

\bibitem{LEN93} G. Lenz {\it et al.}, Phys. Rev. Lett. {\bf 71}, 
3271 (1993).

\bibitem{DEN99} L. Deng {\it et al.}, Nature {\bf 398}, 218 (1999).

\bibitem{JAV86} J. Javanainen, Phys. Rev. Lett. {\bf 57}, 3164
(1986); A. Smerzi {\it et al.}, {\it ibid.} {\bf 79}, 4950 (1997); 
J. Ruostekoski and D.F. Walls, Phys. Rev. A {\bf 56}, 2996 (1997); 
M.W. Jack {\it et al.}, {\it ibid.} {\bf 54}, 
R4625 (1996); I. Zapata {\it et al.}, {\it ibid.} {\bf 57} R28 (1998); 
J. Williams {\it et al.}, {\it ibid.} {\bf 59} R31 (1999)

\bibitem{DAL96} F. Dalfovo and S. Stringari, Phys. Rev. A {\bf 53}, 
2477 (1996); G. Baym and C.J. Pethick, Phys. Rev. Lett. {\bf 76}, 6 (1996);
D.S. Rokshar, {\it ibid.} {\bf 79}, 2164 (1997);
R.J. Dodd {\it et al.}, Phys. Rev. A {\bf 56}, 587 (1997);
J. Javanainen {\it et al.}, {\it ibid.} {\bf 58}, 580 (1998);
T. Isoshima and K. Machida, {\it ibid.} {\bf 59}, 2203 (1999); 
H. Pu {\it et al.}, {\it ibid.} {\bf 59}, 1533 (1999); 
D.A. Butts and D.S. Rokshar, Nature {\bf 397}, 327 (1999);
S. Stringari, Phys. Rev. Lett. {\bf 82}, 4371 (1999).

\bibitem {JAC98} B. Jackson {\it et al.}, Phys. Rev. Lett. {\bf 80}, 
3903 (1998).

\bibitem{BOL98b} E.L. Bolda {\it et al.}, Phys. Rev. Lett. {\bf 81}, 
5477 (1998).

\bibitem{REI97} W.P. Reinhardt and C.W. Clark, J. Phys. B {\bf 30},
785 (1997);  T.F. Scott {\it et al.}, {\it ibid.} {\bf 31},
329 (1998); Th. Busch and J.R. Anglin, e-print cond-mat/9809408.

\bibitem{KIV98} Y.S. Kivshar and B. Luther-Davies, Phys. Rep.
{\bf 298}, 81 (1998), and references therein.

\bibitem{KAP98} A.E. Kaplan {\it et al.}, Physica Scripta T {\bf 76},
93 (1998); I. Marzoli {\it et al.}, Acta Physica Slovaca {\bf 48},
323 (1998).

\bibitem{CHO99} S. Choi {\it et al.}, unpublished.

\bibitem{RUO99d} 
The dynamics of GPE  with attractive interactions generates bright 
solitons consisting of matter-wave packets.
However, for large attractive 
interactions the BEC remains focused in the center of the trap 
and the box boundary do not play an important role.

\bibitem{MAN98} I. Manek {\it et al.}, Optics Comm. {\bf 147}, 67 (1998).

\bibitem{ERN98} U. Ernst {\it et al.}, Europhys. Lett. {\bf 41}, 1
(1998).

\bibitem{RUO97b} J. Ruostekoski {\it et al.}, Phys. Rev. A {\bf 55},
3625 (1997).

\bibitem{STA99} D.M. Stamper-Kurn {\it et al.}, e-print cond-mat/9906035.

\bibitem{HAL98b} D.S. Hall {\it et al.}, Phys. Rev. Lett. {\bf 81},
1543 (1998).

\end{references}
\end{document}